# Optofluidic Temperature and Pressure Measurements with Fiber Bragg Gratings Embedded in Microfluidic Devices

G. A. Cooksey and Z. Ahmed

National Institute of Standards and Technology (NIST), Sensor Science Division
100 Bureau Drive, Gaithersburg, MD 20899
*zeeshan.ahmed@nist.gov*

## ABSTRACT

The integration of photonic sensors into microfluidic devices provides opportunities for dynamic measurement of chemical and physical properties of fluids in very small volumes. We previously reported on the use of commercially available Fiber Bragg Gratings (FBGs) and on-chip silicon waveguides for temperature sensing. In this report, we demonstrate the integration of FBGs into easy-to-fabricate microfluidic devices and report on their sensitivity for temperature and pressure measurement in microliter volumes. These sensors present new routes to measurement in microfluidic applications such as small-volume calorimetry and microflow metrology.

## 1 INTRODUCTION

Microfluidic platforms provide sub-microliter control of fluid transport and mixing of chemicals over a broad range of length and time scales. However, characterization of fluids processed in microfluidic devices is generally performed with macroscale systems. Much effort is now focused on integrating measurement systems that can rapidly interrogate the composition and physical properties of fluids within the microfluidic systems themselves, where rapid sensing and dynamic decisions can be made while fluids are still on the chip.

Photonic sensing technologies are particularly promising for microfluidics because they facilitate a broad array of sensing modalities and are compatible with the scale of microfluidic systems. Fiber Bragg Gratings (FBGs) are commercially available photonic sensing devices that have high sensitivity to changes in temperature and mechanical strain [1]. In addition, research and development of smaller chip-scale technologies, such as ring resonators or photonic crystal cavities, are taking measurements into even smaller spaces and providing opportunities for parallelization and high-throughput multifunctional sensing platforms [2, 3].

Microscale photonic devices are well matched in scale to couple with microfluidics, but as with microelectromechanical systems, there can be challenges integrating different platforms and material types. We have focused on rapid prototyping with tapes and laminates as a simple method to build and test devices without need of sophisticated equipment [4]. Overall, we demonstrate these devices sense small changes in temperature and pressure that present new opportunities for thermodynamic and mechanical interrogation of microfluidic processes.

## 2 EXPERIMENTAL

The optofluidic devices used in this study are easy to fabricate and assemble with off-the-shelf components. Poly(dimethylsiloxane) (PDMS) membranes were purchased (Interstate Specialty Products)[†] or cured from spun-coat PDMS (Sylgard 184, 10:1 base:crosslinker, Dow Corning). Double-sided silicone tape (#96042, 3M™) was used to adhere layers together and was also cut with a plotter razor cutter (FC800, Graphtech) or laser machine (VLS2.30 VersaLASER, Universal Laser Systems) to form microfluidic channels and channels for FBG insertion [4]. Glass slides and acrylic films (McMaster-Carr) were used as substrates and covers, respectively.

Commercially available FBGs with Bragg resonance in the range of 1540 nm to 1560 nm (os1100, Micron Optics) were adhered in microchannels or cured into spin-cast PDMS membranes of approximately the same thickness as the fiber diameter. We have described in detail elsewhere the experimental system to interrogate photonic devices [5, 6]. Briefly, peak resonance of FBGs was determined after scanning a laser over the resonance region and measuring the power of the reflected signal. Fluid temperatures were controlled using a temperature-controlled water bath. Fluid pressure was controlled using the height of water in a reservoir with additional pressure provided by a pressure controller (PSD-15PSIG, Alicat Scientific) over the reservoir.

## 3 RESULTS AND DISCUSSION

A temperature sensing microfluidic device was fabricated by embedding a FBG under a 500 μm wide microfluidic channel (**Figure 1**). The microchannel spanned approximately 8 mm of the 10 mm grating region and contained ≈ 2 μl of fluid above the FBG sensor. We calibrated the temperature and sensitivity of the fiber in the chip by measuring the Bragg resonance after submersing the entire chip in controlled temperature baths from 4 °C to 25 °C. In this temperature range, we determined a resonance shift of 17.9 pm/°C. This value is nearly double the manufacturer specification of 10 pm/°C, which we believe indicates that the FBG is strained due to thermal expansion of the chip. Sensitivity of the FBG to changing temperatures of fluid in the microchannel was tested by flowing ice water into the chip at different flow rates, while the chip was held at 22 °C. Faster flows resulted in more heat transfer to the sensing region of the microchannel. Peak resonances were mapped to the calibrations in order to estimate the mean temperature at the sensor (**Figure 1d**). It is important to point out that the FBG is not fully exposed to the flowing liquid – it is only exposed at the bottom of the fluid channel and is otherwise buffered by the chip. Thus, the fiber measures only a fraction of the difference between the liquid and chip temperatures. Flows of 500 μl/min dropped the mean temperature of the FBG to about 19 °C. We did not test faster flows, but the data indicate that the sensor has not been maximally cooled by the flowing water.

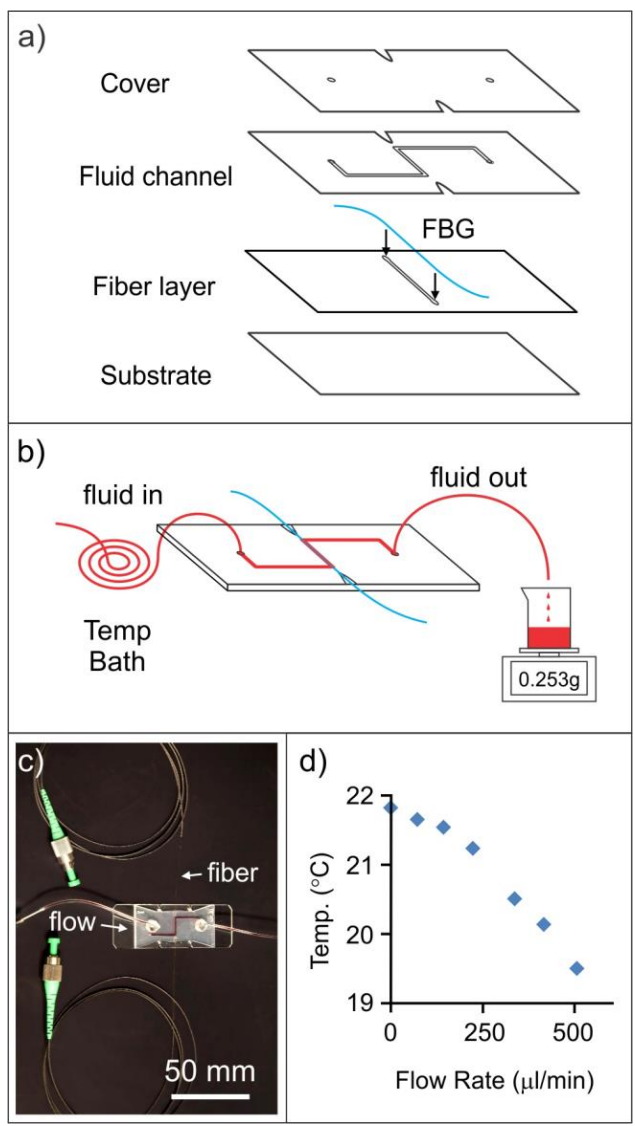

*Figure 1. (a) Exploded view of optofluidic device shows the various layers constructed from laser-cut tapes and laminates. (b) The optofluidic chip layout delivers temperature-controlled liquids over a FBG through a microfluidic channel. Flow was measured by collecting fluid on a microbalance. (c) Photograph of the completed device with FBG and red dye loaded in the microchannel. (d) Heat transfer into the device was controlled by modifying the flow of ice water into the chip. The estimated mean temperature at the FBG is plotted against the flow delivering cold liquid into the 22 °C device.*

Previously, we determined the temperature uncertainty of FBGs in our measurement system was on the order of 0.5 °C [6]. The optofluidic chip demonstrates the capacity to sense perhaps even smaller relative temperature changes, but more work is needed to understand temperature distributions along the microchannel and around the FBG, which was only partially exposed to the cold liquid. It is important to note, however, that compared to macroscale thermometry measurements, the system volume has been reduced to 2 μl above the sensor, which is a reduction of nearly 5 orders of magnitude. Our work indicates an in-channel optical sensor could be used to perform sensitive thermo-chemical measurement, *e.g.* monitoring enthalpic changes due to pseudo-polymorphic transitions in crystalline materials [7].

We initially buried the FBG in a microchannel below the fluid channel as a way to precisely control its position; however, placing the fiber directly in the channel should improve temperature sensitivity by increasing the surface contact with the fiber while reducing the volume flow through the channel. Even smaller channels, nearer to the size of the fiber, could be achieved using higher resolution fabrication strategies, such as soft lithography [8]. Although we controlled heat transfer to the sensor by changing the fluid flow, the device could also be operated as a flow meter by measuring heat transfer and calibrating to determine flow.

Because FBGs are also quite sensitive to mechanical strain, we tested the feasibility of using FBGs as pressure sensors within microfluidic devices. We designed a microfluidic device that incorporated a FBG into a flexible membrane made of PDMS and positioned it below a microchannel. The microchannel above the FBG was expanded to a 2 mm diameter region that could deflect downward into an open region below the membrane (**Figure 2**). Increasing fluid pressure deflected the membrane into open space below the chamber, which strained the fiber and shifted the Bragg wavelength. We tested input fluid pressures ranging from 0 to 75 kPa (0 to 10.9 psi), which cover typical pressures for microfluidic applications [9]. The device was sensitive to pressure changes in the microchannel on the order of about 700 Pa (0.1 psi) although the wavelength shift decreased as pressures eclipsed 5 kPa.

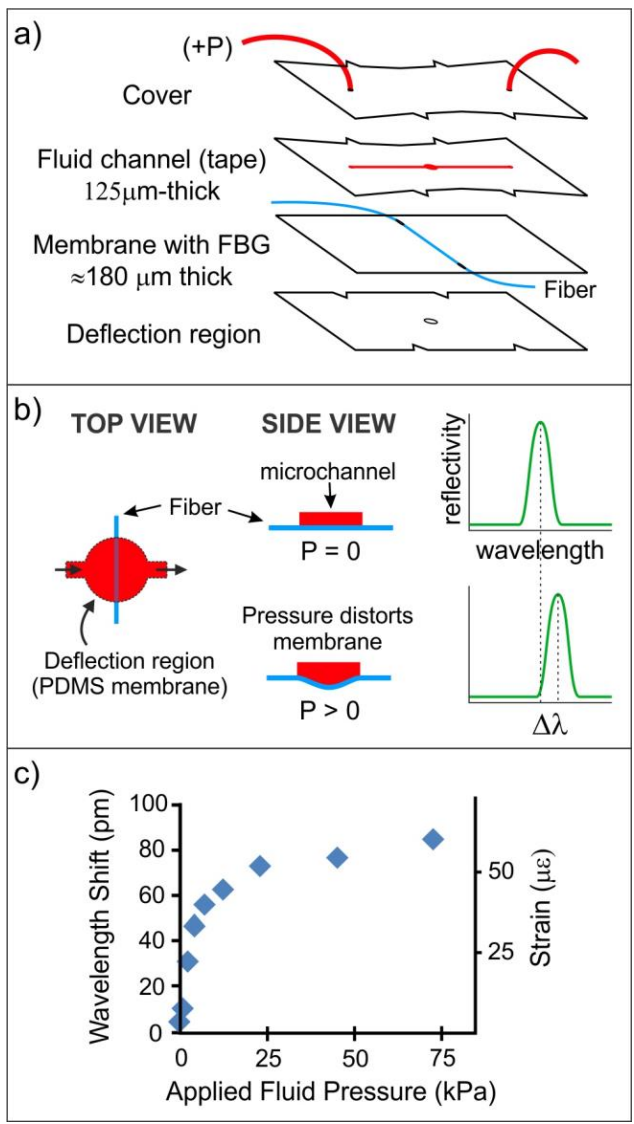

*Figure 2. (a) Exploded view of layers involved in making the pressure sensing microfluidic device. P is applied pressure. (b) An FBG (blue) embedded in a flexible membrane responds to strain applied via a microfluidic channel (red) and deflects down into an open chamber. Bragg resonance shift is depicted in green traces. (c) Bragg resonance peak shift (and equivalent stress) is shown as a function of pressure applied to the input fluid. Wavelength shift is converted to strain based on manufacturer specification of 1.4 pm/µε.*

Further characterization of the pressure sensor is needed. We would like to better understand how membrane deformation relates to the strain on the FBG. It is unclear if the saturation of the Bragg wavelength shift above 25 kPa applied pressure is related to the membrane reaching peak deformation into the deflection region. Because we see the membrane deflecting around the fiber, we are in the process of determining if a more rigid membrane material would expand the dynamic range of the system. We are also looking at the reproducibility of these measurements given that fiber alignment over the deflection region will likely influence sensor performance.

## 4 SUMMARY

This work demonstrates easy-to-build optofluidic systems that enable measurements of small temperature and pressure changes in microliter volumes. These sensing systems provide on-chip assessment of flow and heat transfer, which enables improvement in fluid metrology and permits new opportunities in microscale calorimetry and advanced biological sensing.

## [†]DISCLAIMER